\documentstyle[epsfig,rotate]{mn}
\title{{\it JHK} Standard Stars for Large Telescopes:
the UKIRT Fundamental and Extended Lists.}
\author[T.G.Hawarden et al.]
       {Timothy G. Hawarden$^{1,5}$\thanks{E-mail: tgh@roe.ac.uk},
        S. K. Leggett$^1$, Michael B. Letawsky$^{1,2,3}$, 
        \and David R. Ballantyne$^{1,2,4}$ \& Mark M. Casali$^5$\\ 
        $^1$ Joint Astronomy Centre, 660 N. A`Ohoku Place, Hilo, Hawaii 
        96720, USA\\
        $^2$ Department of Physics and Astronomy, University of Victoria,
        Victoria V8W 3P6, British Columbia, Canada\\
	$^3$ Subaru Telescope, 650 N. A`ohoku Place, Hilo, Hawaii 96720, 
	USA\\
        $^4$ Institute of Astronomy, University of Cambridge, Madingley
        Road, Cambridge CB3 0HA, UK\\
        $^5$ UK Astronomy Technology Centre, Royal Observatory, Blackford
	Hill, Edinburgh EH9 3HJ, UK.}

\date{Accepted 2000 xxxxxx yy.
      Received 2000 xxxxxx yy;
      in original form 2000 xxxxxx yy}
\newcommand{\degree}{\mbox{\,$^\circ$}}

\def\arcsec{\hbox{$^{\prime\prime}$}}
\def\arcmin{\hbox{$^{\prime}$}}

\begin{document}
\maketitle
\begin{abstract}

We present high-precision {\it JHK} photometry with the 3.8m UK Infrared
Telescope (UKIRT) of 83 standard stars, 28 from the widely used
preliminary list known as the ``UKIRT Faint Standards" (Casali \&
Hawarden, 1992), referred to here as the Fundamental List, and 55
additional stars referred to as the Extended List.  The stars have 9.4 $<
K <$ 15.0 and all or most should be readily observable with imaging array
detectors in normal operating modes on telescopes of up to 10m aperture.
Many are accessible from the southern hemisphere. Arcsec-accuracy
positions (J2000, Epoch $\sim$1998) are given, together with optical
photometry and spectral types from the literature, where available, or
inferred from the $J-K$ colour. K-band finding charts are provided for
stars with proper motions exceeding 0.\arcsec3 yr$^{-1}$. We discuss some
pitfalls in the construction of flat fields for array imagers and a method
to avoid them. On 30 nights between late 1994 and early 1998 the stars
from the Fundamental List, which were used as standards for the whole
programme, were observed on an average of 10 nights each, and those from
the Extended List on an average of 6 nights. The average internal standard
error of the mean results for the $K$ magnitudes is 0.$^m$005; for the
$J-H$ colours it is 0.$^m$003 for the Fundamental List stars and 0.$^m$006
for those of the Extended List; for $H-K$ the average is 0.$^m$004. The
results are on the natural system of the IRCAM3 imager, which used a 256 x
256 InSb detector array with ``standard" $JHK$ filters, behind gold-coated
fore-optics and a gold- or silver-dielectric coated dichroic.  We give
colour transformations onto the CIT, Arcetri and LCO/Palomar NICMOS
systems, and preliminary transformations onto the system defined by the
new Mauna Kea Observatory near-infrared filter set.

\end{abstract}


\section{Introduction}

\subsection{Near infrared (NIR) photometric systems}

Definition of a near-infrared photometric system was begun by Johnson and
colleagues ({\it c.f.} Johnson 1966 and references therein). The systems
which eventually came into widespread use fell into three families ({\it
c.f.} Elias {\it et al.} 1983). The South African Astronomical Observatory
(SAAO) system of Glass (1974) is most recently manifest in the work of
Carter (1990) and Carter \& Meadows (1995). The AAO system (Allen \&
Cragg, 1983) derives from this and more especially from the the Mount
Stromlo Observatory (MSO) system ({\it c.f.} Jones \& Hyland, 1982), of
which a more recent descendant is the Mount Stromlo and Siding Springs
Observatory (MSSSO) system ({\it c.f.} McGregor, 1994). The Caltech/Tololo
(CIT) system was first set out by Frogel {\it et al.} (1978) and, in a
wider sample of fainter stars observed to higher precision, in the
defining paper by Elias {\it et al.} (1982). This has become the default
system for NIR photometry in the Northern Hemisphere. Its relation to the
Southern Hemisphere systems was explored by Elias et al. (1983).

Local NIR photometric systems have been developed at the Observatorio
Astron\'{o}mico Nacional, M\'{e}xico (Tapia {\it et al.} 1986, Carrasco
{\it et al.} 1991) and at ESO, where an initial system by Engels {\it et
al.} (1981) was updated by Bouchet {\it et al.} (1991). Not surprisingly
there have been several intercomparisons ({\it e.g.} Elias {\it et al}
1983), sometimes combined with homogenisations (Koorneef 1983a; Bessel \&
Brett 1988).

Absolute calibrations have been undertaken by several methods. Perhaps the
most fundamental approach has been that of the absolute calibration of
Vega in the present wavelength range by comparison with a furnace ({\it
e.g.} Blackwell {\it et al.} 1983). Other approaches have been adopted by,
amongst others, Koorneef {\it et al.}, Bersanelli {\it et al.} 1991 (who
provide a useful summary), Bessel \& Brett (1988) and Carrasco {\it et
al.} (1991). More recently Cohen and colleagues have pursued an
comprehensive programme ({\it c.f.} Cohen {\it et al.} 1999 and references
therein) employing a variety of methods, including in particular
space-based measurements and the use of model atmospheres to link observed
results, to establish a network of bright absolute calibrators over the
sky.

\subsection{UKIRT ``in-house" standards 1980-1992}

At the 3.8m UK Infrared Telescope (UKIRT) an in-house (unpublished) set of
standard stars, largely drawn from the CIT list, was adopted in the early
1980s. Repeated observations of these stars using the UKIRT single-channel
photometer UKT9 and its predecessors were used to accumulate corrections
to the catalogue magnitudes and colours for the individual stars. The list
therefore established a UKIRT ``natural" system, with its $K$ zeropoint
based mainly on the early-type stars of the CIT list, and indeed no
difference could be discerned between the UKIRT and CIT systems at $K$
({\it c.f.} Guanieri, Dixon \& Longmore, 1991). The stars became known as
the UKIRT Standards, and the list was used extensively through the late
1980s and early 1990s to calibrate measurements with the single-channel
``UKT" series of InSb photometers, and also to calibrate spectroscopic
observations.

\subsection{The need for faint NIR standard stars}

Infrared standard stars have traditionally been bright objects, initially,
in part, because early detectors were relatively insensitive. The stars
comprising the standard lists mentioned in the previous section have {\it
K} magnitudes between $\sim$0 and $\sim$8, with the exception of the list
by Carter \& Meadows (1995) which includes about 12 objects which are
fainter than {\it K} $\sim$8.5 and accessible to Northern telescopes.

Modern near-IR array detectors are enormously more sensitive than the
devices used to set up the systems. This sensitivity has been achieved by
dramatic reductions in detector read noise, effectively from reductions in
detector size: maximum sensitivity is achieved by making the detector
capacitance as small as possible, so that signal voltages are maximised
relative to the noise of the readout electronics. The signal current is
measured by discharging the detector capacitance, and as that capacitance
is reduced to increase sensitivity, so the amount of charge that can be
measured in the shortest possible integration in normal observing mode
(the ``well depth") necessarily also reduces. Unless a non-standard
(high-speed) observing mode is employed, such a system effectively has a
fixed dynamic range, and as its sensitivity increases, so the brightest
measurable sources become fainter.

This is well illustrated by developments at UKIRT. After 1985 the IRCAM
series of instruments (McClean {\it et al.} 1986) became the ``workhorse"
imagers. The last of these, IRCAM3 (Puxley {\it et al.}, 1994) was used
from 1991 to 1999 and is the source of all the data presented here.
Because of the abovementioned limitations IRCAM3 could not carry out
broad-band {\it JHK} measurements of stars brighter than {\it K} $\sim$
9.0 in normal observing mode, without saturation, unless the telescope was
defocussed or the seeing very poor. Such constraints will of course be
even more severe for yet larger telescopes.

\section{The UKIRT Faint Standards}

In 1990 a programme was initiated at UKIRT to construct a set of standard
stars faint enough to be observable with the IRCAM imager in standard
observing mode. A list of stars was selected from Landolt's (1983)
equatorial $UBVRI$ standards and the compilation of potential HST
reference sources by Turnshek {\it et al.} 1990, supplemented by
additional main-sequence objects with solar-like colours from the old open
cluster M67 (Eggen \& Sandage, 1964) and subgiants from the globular
clusters M3 and M13, none of which are likely to be variable.

The final list of stars was observed over two years in 1990 and 1991 with
the facility single-channel photometer UKT9, using an InSb detector
employing charge integration on an external capacitance. The observations
were done through apertures of angular diameter 8\arcsec \hspace{0.5mm} or
12\arcsec \hspace{0.5mm} on the sky, with background removal by chopping
with the secondary mirror. The observations were made relative to numerous
stars from the CIT-based UKIRT ``in-house" standards, mentioned above. The
results were made available to users of UKIRT and of other telescopes by
Casali \& Hawarden (1992), in the UKIRT Newsletter. These stars, the
``UKIRT Faint Standards" have been very widely used by observers in the
near-IR on large telescopes.

As acknowledged by Casali \& Hawarden, the internal precision of their
results left something to be desired, especially for the fainter objects.
Also, while the majority of the stars have impeccable photometric
credentials from Landoldt (1983, 1992) some were less well known objects,
for which several observations over a number of years were desirable to
establish constancy. Other drawbacks, such as the paucity of stars both
red enough for transformations and faint enough to observe, and of stars
away from the equator, became apparent. Consequently a further programme
was initiated in 1994, with the intent of strengthening the mutual
precision of the the magnitudes and colours of the UKIRT Faint Standards
and simultaneously of supplementing their number with additional stars,
widely distributed both in location and in colour. These last comprise the
UKIRT Faint Standards Extended List.  Not all of the original UKIRT Faint
Standards have been retained in this version of the Fundamental List: FS
8,9 and 26 (SA94-251, SA94-702 and SA108-475) were omitted because they
were too bright to measure with IRCAM3; FS24 (SA106-1024) is a $\delta$
Scuti variable (Landolt, 1990) which exhibits occasional excursions in the
NIR; FS 25 (SA107-1006) and FS 28 (SA109-71) are double, with separations
that interfere with photometry using aperture radii of a few arcsec. (FS18
is also double but with a small separation.)

The Extended List standards were selected as follows. An initial selection
was made from the Carlsberg Meridian Catalogue (1989), which lists precise
positions and good-quality magnitudes, as well as HD spectral types;
however few CMC stars proved faint enough in the IR.  The majority of the
stars in the Extended List were selected from the Guide Star Photometric
Catalogue (GSPC, Lasker {\it et al.} 1988) with a view to reasonably
uniform sky coverage and inclusion of a range of colours. These were
supplemented by samples of intrinsically red objects (red dwarfs) from 
Leggett \& Hawkins 1988 and Leggett 1992. 

Heavily reddened stars behind dark clouds were also sought, and selected
stars in B216/217 (Goodman {\it et al.} 1992), L1641 (Chen {\it et al}
1993), Serpens/Ophiuchus (Eiroa \& Casali 1992) and Sharpless 106 (Hodapp
\& Rayner, 1991) were included in the observing list. Intrinsic
variability (presumably of young imbedded objects), crowding and problems
with bright, structured backgrounds eliminated all the stars in L1641 and
S106 and several in Serpens. In the final list four stars from B216 and
four from Serpens have been retained.

The B216 objects are designated here by their order in the list of
background (``b") objects in Table 1 of Goodman {\it et al.} (1992). We
note that our $K$ magnitudes agree with the rough NIR results presented by
those authors, but our $J-K$ colours are bluer, by up to several tenths of
a magnitude; the discrepancies do not correlate with colour.

Nine stars in the final list overlap with Persson {\it et al.} 1998;
differences between our natural system and theirs are discussed in \S 8.

The positions of the stars were all checked at the telescope. UKIRT
observations are made after pointing at a ``Nearstar", usually from the
Carlsberg Meridian Catalogue (1989: CMC), which provides positions
accurate to $\sim$ 0.\arcsec1 or better. During this pointing the
relationship of the pointing of the telescope guider to the reference
point on the array is automatically checked and corrected. After the
``Nearstar" pointing the telescope is slewed to the target star, which is
typically less than 2\degree \hspace{0.5mm}away. When this is completed
the telescope pointing gives the position of the star with subarcsecond
precision relative to the CMC star.

Table 1 lists the Faint Standards, giving our identification number,
positions, proper motions (where available from SIMBAD), catalogue
designations, spectral types (from the literature when available, or
inferred from the $J-K$ colour) and optical photometry. References are
given both to the original source from which the star was selected and the
sources of the spectral types and photometry. Identification numbers have
been assigned to the stars as it has been found that short and unambiguous
numbers assist communications in crowded and noisy telescope control
rooms. The 28 stars of the Fundamental List have two-digit identification
numbers while the 55 stars of the Extended List have three-digit numbers.
The positions are accurate to $\sim$1\arcsec \hspace{0.5mm}and are
expressed in the J2000 co-ordinate system, epoch $\sim$1998. Figure 1
gives $K$-band finding charts, epoch 2000.8, for the 10 stars with proper
motions $\geq$0.\arcsec3 yr$^{-1}$. Optical finding charts for most of the
rest may be found in the source references.

Several of the most obscured and intrinsically reddest stars are difficult
or impossible to detect at visible wavelengths, even with large
telescopes, and may therefore require to be observed with offset- (or no)
guiding.

\section{Observations}

\subsection{The instrument: IRCAM3, its fore-optics and filters.}

By 1994 UKIRT was no longer equipped with a photometer, so the array
imager IRCAM3 (Puxley {\it et al.} 1994 and references therein) was
employed for all of these observations. As noted above, stars brighter
than {\it K} $\sim$9.0 were unobservable with IRCAM3 using the normal
readout mode NDSTARE (Chapman {\it et al.} 1990). At that time there were
no standard stars on other systems to which the new observations could be
connected which were faint enough to be observable with IRCAM3 in this
mode. Consequently the new data are entirely self-referential ({\it i.e.}
calibrated entirely relative to the UKIRT Fundamental List Faint Standards
themselves) and are on the natural system of the IRCAM3 imager and its
predecessor. However Guarneri {\it et al.} (1991) showed that this system
is close to that of UKT9, the single channel cryostat with which the
original observations were made and which had conventionally defined the
UKIRT photometric system, and for which transformations to, {\it e.g.},
the CIT system, {\it are} known (Casali \& Hawarden, 1992).

IRCAM3 employed an InSb array of 256x256 pixels, each of which subtended
an angle of 0.\arcsec28 on the sky. The optical system included three
lenses, two of BaF$_2$ and one of LiF, with total attenuation factor of
$\sim$ 0.79, imperceptibly dependent on wavelength across the range of
interest here. It used gold-coated reflecting fore-optics (2 surfaces) and
received the infrared beam by reflection from a tertiary mirror carrying a
dichroic coating. Until the end of 1996 this was of gold, $\sim$40 nm
thick, with $\sim$25\% transmittance for green light, which allowed source
acquisition, and guiding, in the optical. The gold coatings (on both
dichroic and fore-optics) have proved very stable over the years and no
evolution of the UKIRT colour system was ever detected.

The observations were made with the ``standard" UKIRT broad-band filters:
$J$ (1.28$\mu$m), $H$ (1.65$\mu$m), and $K$ (2.20$\mu$m), transmittance
curves for which are specified in Table 2.  We also list in
the Table the effective transmittance derived by convolving the profiles
with a calculated atmospheric transmission for Mauna Kea assuming 1.2~mm
of precipitable water.  The $J$ and $H$ filters were manufactured by Barr
Associates, the $K$ filter by Optical Coating Laboratories Inc. (OCLI).

As part of a programme of telescope enhancements, at the end of 1996 the
gold dichroic coating was replaced with a proprietary silver-dielectric
(``Ag-di") multilayer coating with similar reflectivity in the IR but much
better transmission at visible wavelengths. Nominal reflectance curves for
the gold fore-optics (two reflections) and for the gold and ``Ag-di"
dichroic coatings are given in Table 3. For the gold coatings these are
generic curves, while that for the silver-dielectric dichroic coating is
from a witness sample. It has not been possible to measure the actual
dichroics because of the large size of the glass substrates.

Examination of Tables 2 and 3 indicates that the change in the effective
wavelength of the $J$ filter (the only one likely to be affected by the
change from gold to Ag-di dichroic coating) should be well under 1\%, so
it is not surprising that observations of stars with a wide range of
colours made before and after the change in dichroic coating revealed no
discernable change in the $J$-band system properties. However changes in
{\it throughput} (zeropoint) with aging of the dichroic coating {\it have}
been observed in the $J$ and other bands.

IRCAM3 has now been modified by the addition of internal, cold, magnifying
optics, giving a much finer pixel scale and eliminating the warm gold
fore-optics. The instrument with which the present observations were
caried out is therefore no longer in existence in its original form.

\subsection{Observing procedures}

The NDSTARE non-destructive ``up-the-ramp" array readout mode (Chapman
{\it et al.} 1990) was employed for all observations in this programme.

The array field-of-view was $\sim$72\arcsec. A series of frames were
taken, with the star image located first in the middle of the array and
then on a grid of 2 (later 4) other positions, offset by $\pm$8\arcsec
\hspace{0.5mm}in RA and in DEC. The use of several offset exposures allows
correction for the effects of defective pixels.

The resulting images were later combined to build a mosaic, one
through each filter, as described in the next section. Exposures varied
from 0.5 to 8 seconds between array readouts (``on-chip" exposure times).
When exposures shorter than 5 seconds were employed, several
images were coadded in the readout electronics to accumulate at least 5
seconds total exposure before moving to the next array position.

Initially only three images were taken for each mosaic, in order to
minimise overheads associated with the crosshead movement and
re-acquisition of the images by the autoguider. This proved to have
serious drawbacks for the data reduction (see below) so the number of
images per mosaic was increased to five and the individual exposure times
reduced to give the same total exposure. The observing algorithms
(``EXEC"s) were adjusted to give a nominal signal-to-noise ratio (S/N)
$\geq$100 for $K <12$ and $\geq$ 30 for $K \sim15$.

As noted above, the observations were carried out using the stars of the
Fundamental List as standards. As many as possible of these were observed
on each of the 30 nights or part nights devoted to this programme. An
average of 10 (minimum 5, maximum 21) Fundamental List stars were observed
on each night.

\section{Image Data Reduction}

Linearity corrections were applied to the raw IRCAM data. Dark frames, of
the same duration as the on-chip exposure times employed for the
observations, were then subtracted from the images and the result scaled
to one second equivalent on-chip exposure time. The images were then
normalised to the same average sky signal.

In array photometry ``flat-field" frames are required to adjust the
sensitivities of the pixels to a uniform value over the array. This is
generally done by taking exposures of a target area on the inside of the
dome. 

For reasons which are not well understood, satisfactory images for
constructing flat-field frames from dome targets have never been
successfully secured with imagers at UKIRT. This has necessitated a
reliance on images of the sky. A number of exposures, long enough to
have a strong background signal on each pixel, are used to determine a
flat field frame by median filtering, {\it i.e.} to form a frame in
which the signal in each pixel is the median of the signals in that
pixel in the input images.

Because significant instabilities were known to affect the detector array
it was believed that flat field corrections should be redetermined at
frequent intervals. In the interests of efficiency, the frames secured in
the current mosaic were employed. However the use of small numbers of
frames leaves the resulting median subject to a potentially important
bias, discussed below, and in retrospect it would have been preferable to
determine flat-field corrections from occasional sets of deeper and more
numerous exposures taken at intervals through each period of observing.

Each image was divided by the appropriate flat-field frame (see below)
to remove sensitivity variations. 

The final stage of the preparation of the observations for photometric
analysis was to combine the three or five reduced frames into a single
mosaic frame by adding them together, displaced by the known shifts of the
crosshead or telescope, or (optimally) so as to bring the star images on
the individual frames into alignment. Since the array has an effective
filling factor of ~100\%, sampling and spatial resolution were not
significant issues for the photometry, despite the large pixel size, and
the precision of this alignment was not critical.

\section{Flat-field bias}

\subsection{Errors of the median}

The median of an odd numbered set of data values is the value at the
centre of the distribution of values; for an even numbered set of values
it is the mean of the two values closest to the centre. If one element of
the data set is in fact not part of the true distribution ({\it i.e.} is
spuriously too low or too high) then ignoring this fact is
effectively to use the incorrect algorithm. This will produce a
result which is in error by half the difference of the central values of
the distribution; the sign of the error will be positive ({\it i.e.} the
result will be too large) if the spurious datum is itself at the high
end of the distribution, and negative if the spurious value is low.

In the present context, a pixel with the target star on or near it in a
particular frame is {\it certain} to have a high signal. Clearly the
proper data set from which to determine the median {\it background} level
in that pixel should exclude the frame with the star. However, if this
frame is retained, the apparent median for the affected pixel(s) will, by
the argument above, be spuriously increased by an amount $\delta$ equal to
half the difference of the signal level values adjacent to the median in
that pixel.

\subsection{Effects of median errors}

In the flat field frame that pixel would therefore be set so as to {\it
reduce} the signal by the corresponding amount in processing an image
frame. This is true of all the pixels underlying the particular star
image. If the resulting flat field is then used in the reduction of the
image which {\it caused} the problem, the signal level in both the star
image and underlying background will be spuriously reduced. The reduction
factor for a given pixel with median signal {\it m} is therefore  

\begin{displaymath}
R = 1 - \frac{\delta}{m}
\end{displaymath}

If there are few (N) frames $\delta$ is of the order of the average
pixel-to-pixel difference of the background measured on those frames
divided by $\sqrt{N}$.

Flat field bias of this type can be seen in the flat field frame as
holes where stars appeared in the input frames (these are all
approximately the same depth irrespective of the brightess of the
originating star).

The effect of the bias hole will be to reduce the signal from the star
{\it and underlying background} over the whole area of the star image. The
brightness of a source is measured by subtraction of a general sky
background from the total signal in an aperture surrounding the star. For
bright objects, for which the background is negligible, the effect of the
flat-field hole will therefore just be a scaling of the source signal
by the flat-field bias, i.e. the star will appear too faint by that small
fraction. However for fainter objects the background underlying the source
image becomes a significant part of the signal from the star aperture;
because it is scaled too, the result is a deficit in the total ``star"
signal relative to the sky level which is subtracted to determine the
source flux. Fainter sources will be progressively more seriously
underestimated.

\subsection{Avoiding flat-field bias}

For imaging of {\it target} objects at known positions on each frame (such
as standard stars) the bias is readily eliminated by deriving a separate
flat field frame for each image of a mosaic set using only the {\it other}
frames of the set, on which the target is guaranteed to be somewhere other
than where it is on the particular frame being reduced.

However this solution meant that the individual early three-frame mosaic
observations were reduced using flat fields derived from only two images.
The ``median" then derived is in fact the mean of the sky signals in each
image: and an average of only two frames is drastically affected if a
pixel in one frame has some other (non-target) star in it. Under these
circumstances an interloping star changes the flat field by half the
star's signal, producing a spectacular hole in the reduced image. In the
present programme, if such holes occurred in the measuring aperture all
three-frame observations of the field were discarded.

\section{Photometry}

The individual reduced frames were automatically searched for the star
image by identifying the brightest pixel in a 20\arcsec \hspace{0.5mm} box
centred on the expected location in the frame of the image of the star.

Automated aperture photometry was then performed on both the individual
reduced images as well as the mosaic image. In all cases a numerical
object aperture 8\arcsec \hspace{0.5mm} in diameter and a concentric sky
annulus with inner diameter 12\arcsec \hspace{0.5mm} and outer diameter
16\arcsec \hspace{0.5mm} were used. The median signal in the sky annulus
(effectively immune to the bias discussed above from the inclusion of
stars or holes, because values from hundreds of pixels were involved in
its derivation) was taken as the sky signal per pixel to be subtracted
from the object counts to get the true star signal, which was converted to
an instrumental magnitude in the usual way.

For the stars of the Fundamental List the instrumental magnitude and the
current catalogue magnitude were added to produce a zeropoint magnitude
(conventionally the magnitude of a source giving one data number per
second through the telescope and photometer system).

\subsection{Identifying faulty images}

During the initial extraction of the photometric data the images were
displayed one by one as they were analysed. This allowed poor images to be
noted and possibly discarded. During the subsequent data analysis the
standard deviation ($\sigma$) of the zeropoints of the individual images
of the mosaic were calculated.  If this standard deviation exceeded
0.$^m$1 then that image was flagged and re-checked visually. If the
auto-selection for that image was incorrect, then the star was selected
manually.  If the image was found to be too distorted to measure ({\it
e.g.} because of bad tracking) it was discarded. If on visual inspection
no problems were evident the photometry of the flagged images was retained
in the data set.

\subsection{Atmospheric extinction and system zeropoints}

The zeropoints from each mosaic were differenced to form raw $J-H$ and
$H-K$ colours as well as $K$ magnitudes, which were plotted against
airmass to derive extinctions. Reductions were performed using the colours
rather than the magnitudes in order to preserve the small but significant
gain in precision that arises when different wavelengths are observed
consecutively in time with minimum change to the instrument, {\it i.e.}
when {\it the differential changes} are minimised.

Linear fits to the plots against airmass were used to determine extinction
coefficients and overall (zenith) zeropoints in each magnitude and colour
for each night. The measurements were reduced to the zenith, not to
outside the atmosphere, to preserve their internal precision. Absolute
precision is much harder to achieve because of strong non-linearity of
extinction curves between 1.0 and zero airmass arising from the presence
of saturated absorption features (Forbes, 1848; for more accessible
references and discussion see Young, Milone \& Stagg, 1994). The
extinction curves between 2.0 and 1.0 airmass are also in fact non-linear,
but the departure from linearity is expected to be unimportant for
reductions to the zenith ({\it c.f.} Young {\it et al.} 1994).

The data were filtered using the dispersions of the individual
frames about the mosaic mean: magnitudes or colours with quadratically
combined errors $\sigma$ $>$0.$^m$1, were omitted from the analysis. The
zeropoints and extinction coefficients were in turn used to reduce the
measurements of the stars on the Extended List. For the stars of the
Fundamental List the residuals of all measurements from the extinction
curves were ledgered.

\subsection{Iterative error trapping}

The results for each star on each night were tabulated and overall means
and standard deviations $\sigma$ of $J-H$, $H-K$, and
$K$ determined.

An iterative error trapping process was then applied: the measurement with
the largest residual in each set of results was identified and new means
and $\sigma$s calculated with that measurement omitted; if thereafter its
residual was more than 3$\sigma$ from the new mean, the images used in
producing the flagged results were once again re-examined for problems
(stars or ``holes" in the measuring aperture; poor images): if any were
found the observation was re-reduced without the offending frame(s), or
rejected.

The nightly extinction curves and zeropoints were then redetermined using
the new results. The re-reduced colours and magnitudes were once again
tabulated and the above procedure repeated until the images associated
with all flagged residuals were double-checked and found to be
satisfactory.  

A final error trapping iteration was then carried out on the finished
data table, using the same 3$\sigma$ criterion as before. This time a
decision was taken whether to include or exclude a measurement from the
final data set, based for example on the now well-known ``typical" error
of an observation: {\it e.g.} measurements differing from the mean by
more than 3$\sigma$, simply because that particular data set had an
atypically small dispersion, were identified and retained.

\section{Results}

Table 4 gives FS (Faint Standard) numbers, positions, the mean {\it K}
magnitudes, $J-H$ and $H-K$ colours and the standard errors of these means
(SEM), together with the number of nights (N) of observations contributing
to the final results. Note that since the stars of the Fundamental List
were the reference stars for the total observing list, these stars were
generally observed several times more often than the Extended List stars.
For the Fundamental List the mean SEM is 0.$^m$0045 for $K$, 0.$^m$0034
for $J-H$ and 0.$^m$0040 for $H-K$, and the stars were observed between 10
and 31 times on an average of 10 nights. Each star of the Extended List
was observed once or (occasionally) twice on each of an average of 6
nights. The average internal standard error of the mean results are
0.$^m$005 for the $K$ magnitudes, 0.$^m$006 for the $J-H$, and 0.$^m$004
for the $H-K$ colours. (The larger average $J-H$ residual for the Extended
List arises from the inclusion of very red objects which are markedly
fainter at $J$.)

\section{Comparisons with Other Systems}

\subsection{The CIT system}

Casali \& Hawarden (1992) list transformations from the UKT9 natural
system at UKIRT to the CIT system:\\
\\
\noindent
$K_{CIT}$ = $K_{UKIRT}$ - 0.018$(J - K)_{UKIRT}$\\
$(J - K)_{CIT}$ = 0.936$(J - K)_{UKIRT}$\\
$(H - K)_{CIT}$ = 0.960$(H - K)_{UKIRT}$\\
$(J - H)_{CIT}$ = 0.920$(J - H)_{UKIRT}$\\

As noted above Guarnieri {\it et al.} 1991 found that differences between
UKT9 and IRCAM photometry at UKIRT must be less than 2\% for stars with
$J - K < 1$ and so the above transformations are valid for converting
the results presented here to the CIT system.

\subsection{LCO/Palomar NICMOS}

Persson {\it et al.} (1998) present $JHK$ and $Ks$ results for 65
solar-type stars with 10.$^m$4 $<$ K $<$ 12.$^m$2, covering both
hemisphere of the sky, together with 27 redder objects to assist in
determining system transformations. The internal precision appears
to be slightly poorer than the current results.

The UKIRT Extended List contains 9 stars in common with Persson {\it et
al.}.  Eight of these (P525-E, P533-D, P161-D, P212-C, P266-C, S867-V,
P565-C and S893-D) are solar-type and one (LHS2397a) is a red dwarf, found
to be slightly variable at I by Martin, Zapatero-Osorio \& Rebolo (1996).
There are no obvious colour terms seen in this limited comparison sample
but there does seem to be a constant zeropoint difference such that
(removing 1-2 outliers for each mean and standard deviation):\\
\\ 
\noindent 
$J_{NICMOS} - J_{UKIRT} = 0.034 \pm 0.004$\\ 
$H_{NICMOS} - H_{UKIRT} = 0.027 \pm 0.007$\\ 
$K_{NICMOS} - K_{UKIRT} = 0.015 \pm 0.007$\\

\subsection{Arcetri/ARNICA system}

Hunt {\it et al.} (1998) observed fields around the northern 22 of the
original UKIRT Faint Standards, as well as 15 A0 stars selected from the
SAO catalogue and 3 deep CCD fields. These 40 fields contain 86 stars
measured on between 3 and 10 nights and which have 7.$^m$9$ < K <
$14.$^m$5. These data are also calibrated relative to the UKIRT Faint
Standards. Their internal precision appears to be similar to that of the
present results except for the fainter objects: the two data sets agree
within their combined errors except for the 4 stars with $K >13.5$, which
are all discrepant by 2$\sigma$ or more.

\subsection{Mauna Kea Observatory near-infrared (MKO-NIR) system}

Both UKIRT cameras (the UKIRT Fast Track Imager, UFTI, and the modified
IRCAM) are now equipped with filters from the Mauna Kea Consortium (the
Mauna Kea Observatory Near-IR (MKO-NIR) filter set: see Simons \&
Tokunaga, 2001 and Tokunaga \& Simons 2001).

These filters have been designed for maximum throughput and minimum
background (and therefore maximum sensitivity) at the same time as having
a much better match to the atmospheric windows than their precursors. This
last implies minimal sensitivity of the results to atmospheric water
vapour (the absorption spectrum of which defines most of the natural
NIR atmospheric windows). Because this excludes most of the saturated
absorption features which give rise to the Forbes Effect (c.f. Young {\it
et al.}, 1994, and references therein) the extinction curves for these
filters are expected to show little or no non-linearity, even between 1.0
and zero airmass. Comparison of results between sites (and nights) with
differing water vapour columns should be greatly facilitated thereby, as
will the determination of extra-atmospheric (absolute) fluxes. The
filters are also of excellent optical quality.

The MKO-NIR system has been endorsed by the IAU Working Group on Infrared
Photometry as the preferred standard photometric system for the near
infrared.

An MKO-NIR filter set has been in UFTI since commissioning in October
1998, and replaced the old $JHK$ set in IRCAM prior to commissioning of
its new plate scale in September 1999. Both the $J$ and $H$ filters are
significantly different from the old filters and hence the magnitudes for
the Faint Standards on the old IRCAM3 system are no longer on the natural
systems of the imagers. Colour transformations have been derived in two
ways: one empirically based on UFTI photometry and the other calculated by
convolving the known filter profiles with spectroscopic data for a
representative set of red stars. The two determinations agree well. Note
that the transformations for IRCAM after its modifications should be
identical with those for UFTI, since its optics have flat transmission
curves across the $J$, $H$ and $K$ windows and both array response curves
are also flat across this wavelength range.

The transformations between the old IRCAM3 system and the new MKO-NIR
system at $H$ and $K$ are well behaved and single-valued. However for the
$J$ filter different terms have to be applied depending on whether or not
the standard star has intrinsic water absorption features. This is due to
the fact that the MKO-NIR $J$ filter cuts off at a shorter wavelength than
the old filter, specifically to avoid water absorption in the terrestrial
atmosphere. We have determined that for stars with no intrinsic water
features the colour transformations between the MKO-NIR system and the
system of the results presented here are:\\

\noindent
$        K_{UKIRT} = K_{MKO} + 0.020[+/-0.005](J-K)_{MKO}$ \\
$      (J-H)_{UKIRT} = 1.040[+/-0.010](J-H)_{MKO}$ \\
$        (H-K)_{UKIRT} = 0.830[+/-0.010](H-K)_{MKO}$ \\
$          (J-K)_{UKIRT} = 0.960[+/-0.010](J-K)_{MKO}$\\
\\
or:\\

\noindent
$   K_{MKO} = K_{UKIRT} - 0.020[+/-0.005](J-K)_{UKIRT}$ \\
$        (J-H)_{MKO} = 0.960[+/-0.010](J-H)_{UKIRT}$ \\
$             (H-K)_{MKO} = 1.205[+/-0.010](H-K)_{UKIRT}$ \\
$                  (J-K)_{MKO} = 1.040[+/-0.010](J-K)_{UKIRT}$ \\
\\

However for stars with significant water absorption, {\it i.e.}stars of
spectral types M4 through late L (but not including the T class
with methane absorption): \\

\noindent
$   K_{MKO} = K - 0.020[+/-0.005](J-K)_{UKIRT}$ \\
$            (J-H)_{MKO} = 0.870[+/-0.010](J-H)_{UKIRT}$ \\
$               (H-K)_{MKO} = 1.205[+/-0.010](H-K)_{UKIRT}$ \\
$               (J-K)_{MKO} = 0.980[+/-0.010](J-K)_{UKIRT}$ \\
\\

We expect soon to re-observe all the UKIRT Faint Standards, and all
accessible LCO/Palomar NICMOS standards, using the new filter set, in
order to place these stars accurately onto the MKO-NIR system. 

\subsection*{Acknowledgments}

This programme was made possible by the use of ``contingency"
engineering time during the UKIRT Upgrades Programme, 1993-1998 ({\it
c.f.} Hawarden {\it et al.} 1999). We are therefore especially grateful
to Dr. Tom Geballe, then Head of UKIRT Operations, for assisting us to
exploit this opportunity. Many of our other colleagues at UKIRT provided 
extensive advice, assistance and support. We are grateful to Prof. Marcia
Rieke for providing us with a preliminary list of the NICMOS stars which
enabled us to incorporate a subset in the present programme.


\newpage
\begin{figure}

\epsfig{file=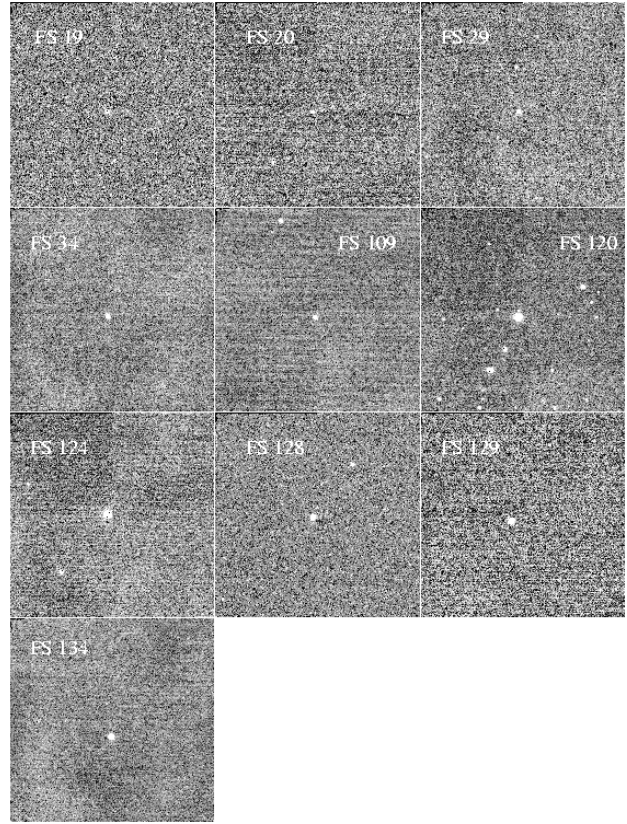,height=12cm}

\caption[]{Finding charts (epoch 2000.8) for the UKIRT Faint Standards
with proper motions larger than 0.\arcsec3 yr$^{-1}$. The images were
taken though a $K$ filter. Each is 93\arcsec \hspace{0.5mm} square; North
is up, East to the left.}

\end{figure}

\newpage
\begin{table*}

\caption[]{The UKIRT Faint Standards: names, spectral types, optical photometry and references.}

\begin{flushleft}

\begin{tabular} {rrrrrlccrrrrr}

FS & \multicolumn{2}{c}{RA \hspace{2mm}(J2000) \hspace{2mm} Dec} & PM-RA  & PM-Dec & Other & Source & Spectral & Spec.
&\multicolumn{1}{c}{$V$} &\multicolumn{1}{c}{$B-V$} &\multicolumn{1}{c}{$V-I$} & Phot.\\
    &             &             &\arcsec \hspace{0.5mm}yr$^{-1}$&\arcsec \hspace{0.5mm}yr$^{-1}$& Name  & Ref.&
Type  & Ref.&       &      &      & Ref.\\
    &             &             &        &        &           &     &       &    &             &  \\
101 & 00 13 43.58 & +30 37 59.9 &$-$0.005&$-$0.009& CMC 400101& 11  & F0    & 21 & 11.7: &  0.0: &    -  & 11\\
102 & 00 24 28.50 & +07 49 00.1 &      - &      - & P525-E    & 12  & G3    & *  & 13.90 &  1.17 &    -  & 12\\
  1 & 00 33 54.48 &$-$12 07 58.1&  0.152 &$-$0.179& G158-100  & 2   & DK-G  & 2  & 14.89 &  0.69 &    -  &  2\\
103 & 00 36 29.60 & +37 42 54.3 &      - &      - & P241-G    & 12  & K2    & *  & 14.32 &  1.05 &    -  & 12\\
  2 & 00 55 09.93 & +00 43 13.1 &     -  &      - & SA92-342  & 1   & F5    & 5  & 11.61 &  0.44 &  0.54 &  9\\
  3 & 01 04 21.63 & +04 13 36.0 &     -  &      - & Feige 11  & 1,2 & sdB   & 6  & 12.06 &$-$0.24&$-$0.26&  9\\
104 & 01 04 59.43 & +41 06 30.8 &  0.000 &$-$0.004& P194-R    & 12  & A7    & *  & 11.03 &  0.18 &    -  & 12\\
105 & 01 19 08.19 & +07 34 11.5 &      - &      - & P527-F    & 12  & K1    & *  & 13.43 &  1.03 &    -  & 12\\
106 & 01 49 46.94 & +48 37 53.2 &      - &      - & P152-F    & 12  & K4    & *  & 14.77 &  1.31 &    -  & 12\\
107 & 01 54 10.14 & +45 50 38.0 &$-$0.025&$-$0.004& CMC 600954& 11  & G0    & 21 & 11.3: &  0.7: &    -  & 11\\
    &             &             &        &        &           &     &       &        &           &  \\
  5 & 01 54 34.65 &$-$06 46 00.4&      - &      - & Feige 16  & 1,2 & A0    & 6  & 12.41 &$-$0.01&$-$0.00&  9\\
  4 & 01 54 37.70 & +00 43 00.5 &      - &      - & SA93-317  & 1   & F5    & 5  & 11.55 &  0.49 &  0.59 &  9\\
  6 & 02 30 16.64 & +05 15 51.1 &  0.071 &$-$0.025& Feige 22  & 1,2 & DA3   & 8  & 12.80 &$-$0.05&$-$0.21&  9\\
  7 & 02 57 21.21 & +00 18 38.2 &      - &      - & SA94-242  & 1   & A2    & 5  & 11.73 &  0.30 &  0.36 &  9\\
108 & 03 01 09.85 & +46 58 47.7 &  0.001 &$-$0.001& CMC 502032& 11  & F8    & 21 & 11.2: &  0.4: &  -    & 11\\
109 & 03 13 24.16 & +18 49 38.4 &  1.246 &$-$1.087& LHS 169   & 13  & M2V   & 14 & 14.13 &  1.45 &  1.72 & 14\\
110 & 03 41 02.22 & +06 56 15.9 &      - &      - & P533-D    & 12  & G5    & *  & 13.20 &  0.70 &    -  & 12\\
111 & 03 41 08.55 & +33 09 35.5 &  0.003 &  0.003 & CMC 601790& 11  & G5    & 21 & 11.3: &  0.9: &    -  & 11\\
112 & 03 47 40.70 &$-$15 13 14.4&      - &      - & S618-D    & 12  & G0    & *  & 12.30 &  0.55 &    -  & 12\\
 10 & 03 48 50.20 &$-$00 58 31.2&      - &      - & GD50      & 2   & DA2   & 8  & 13.99 &$-$0.21&    -  &  8\\
    &             &             &        &        &           &     &       &    &    &     &    & \\
113 & 04 00 14.07 & +53 10 38.5 &      - &      - & P117-F    & 12  & K0    & *  & 14.90 &  1.03 &    -  & 12\\
114 & 04 19 41.72 & +16 45 22.4 &      - &      - & Hy214     & 15  & M7V   & 15 & 21.05 &   -   &  4.17 & 16\\  
115 & 04 23 18.17 & +26 41 16.4 &      - &      - & B216-b5   & 18  & -     & -  &   -   &   -   &   -   & - \\
116 & 04 23 50.18 & +26 40 07.7 &      - &      - & B216-b7   & 18  & -     & -  &   -   &   -   &   -   & - \\
117 & 04 23 56.61 & +26 36 38.0 &      - &      - & B216-b9   & 18  & -     & -  &   -   &   -   &   -   & - \\
118 & 04 24 33.49 & +26 33 37.8 &      - &      - & B216-b13  & 18  & -     & -  &   -   &   -   &   -   & - \\
 11 & 04 52 58.92 &$-$00 14 41.6&      - &      - & SA96-83   & 1   & A3    & 5  & 11.72 &  0.18 &  0.19 &  9\\
119 & 05 02 57.44 &$-$01 46 42.6&  0.001 &$-$0.006& SAO 131719& 11  & A2    & 21 & 10.14 &  0.10 &   -   & 11\\
 12 & 05 52 27.66 & +15 53 14.3 &  0.096 &$-$0.189& GD71      & 1   & sdO   & 8  & 13.03 &$-$0.25&  0.30 &  9\\
 13 & 05 57 07.59 & +00 01 11.4 &      - &      - &  SA97-249 & 1   & G4V   & 5  & 11.74 &  0.65 &  0.72 &  9\\
    &             &             &        &        &           &     &       &    &   &     &     & \\
120 & 06 14 01.44 & +15 09 58.3 &  0.751 &$-$1.220& LHS 216   & 13  & M1V   & 14 & 14.66 &  1.62 &  2.08 & 14\\
121 & 06 59 46.82 &$-$04 54 33.2&      - &      - & S772-G    & 12  & K3    & *  & 14.21 &  1.20 &   -   & 12\\
122 & 07 00 52.02 & +48 29 24.0 &      - &      - & P161-D    & 12  & G0    & *  & 12.85 &  0.57 &   -   & 12\\
 14 & 07 24 14.40 &$-$00 33 04.1&      - &      - & Rubin 149 & 2   & O9-B2p& 2  & 13.86 &$-$0.14&   -   &  2\\
 15 & 08 51 05.81 & +11 43 46.9 &      - &      - & M67-I-48  & 3   & G5IV-V& *  & 14.05 &  0.70 &   -   &  3\\
123 & 08 51 11.88 & +11 45 21.5 &$-$0.008&$-$0.006&P486-R (Note 4)&12& B8V  & 19 & 10.02 &$-$0.08&$-$0.07& 20\\
 16 & 08 51 15.01 & +11 49 21.2 &      - &      - & M67-IV-8  & 3   & G1V   & *  & 14.18 &  0.61 &   -   &  3\\
 17 & 08 51 19.31 & +11 52 10.4 &      - &      - & M67-IV-27 & 3   & G4V   & *  & 13.95 &  0.61 &   -   &  3\\
 18 & 08 53 35.51 &$-$00 36 41.7&      - &      - & SA100-280 & 1   & F8    & 5  & 11.80 &  0.49 &  0.59 &  9\\
124 & 08 54 12.07 &$-$08 04 58.9&  0.939 &$-$0.810& LHS 254   & 13  & M5V   & 14 & 17.41 &  1.75 &  3.97 & 14\\
    &             &             &        &        &           &     &       &    &       &       &       &   \\
125 & 09 03 20.60 & +34 21 03.9 &      - &      - & P259-C    & 12  & G8    & *  & 12.11 &  0.67 &   -   & 12\\
126 & 09 19 18.73 & +10 55 54.2 &      - &      - & P487-F    & 12  & K3    & *  & 14.48 &  1.03 &   -   & 12\\
127 & 10 06 29.03 & +41 01 26.6 &      - &      - & P212-C    & 12  & F9    & *  & 13.03 &  0.52 &   -   & 12\\
 19 & 10 33 42.75 &$-$11 41 38.3&$-$0.297&$-$0.055& G162-66   & 1,2 & DA2   & 8  & 13.01 &$-$0.16&$-$0.27&  9\\
128 & 11 05 10.40 & +07 06 48.7 &$-$0.508&$-$0.156& LHS 2347  & 13  & M5V   & 14 & 19.00 &   -   &$-$3.88& 14\\
 20 & 11 07 59.93 &$-$05 09 26.1&$-$0.038&$-$0.426& G163-50   & 1,2 & DA3   & 8  & 13.06 &  0.04 &$-$0.16&  9\\
129 & 11 21 48.95 &$-$13 13 07.9&$-$0.400&$-$0.349&LHS 2397a&13 (Note 1)&M8V& 14 & 19.57 &   -   &  4.62 & 14\\
130 & 11 24 55.92 & +34 44 38.5 &      - &      - & P264-F    & 12  & K4    & *  & 15.07 &  1.40 &   -   & 12\\
 21 & 11 37 05.15 & +29 47 58.4 &$-$0.147&$-$0.006& GD140     & 4   & DA3   & 8  & 12.50 &$-$0.06&   -   &  8\\
131 & 12 14 25.40 & +35 35 55.6 &      - &      - & P266-C    & 12  & F8    & *  & 12.56 &  0.51 &   -   & 12\\

\end{tabular}

\end{flushleft}

\end{table*}

\begin{table*}

\contcaption{}

\begin{flushleft}

\begin{tabular} {rrrrrlccrrrrr}

FS & \multicolumn{2}{c}{RA \hspace{2mm}(J2000) \hspace{2mm} Dec} & PM-RA  & PM-Dec   
& Other & Source & Spectral & Sp. &\multicolumn{1}{c}{$V$} &\multicolumn{1}{c}{$B-V$} 
&\multicolumn{1}{c}{$V-I$} & Phot.\\
    &             &             & \arcsec \hspace{0.5mm}yr$^{-1}$& \arcsec
\hspace{0.5mm}yr$^{-1}$& Name & Ref.& Type  & Ref.& & & & Ref. \\
    &             &             &        &        &           &     &       &    &       &       &       &   \\
132 & 12 21 39.36 &$-$00 07 13.3&      - &      - & S860-D    & 12  & G1    & *  & 13.33 &  0.55 &    -  & 12\\
 33 & 12 57 02.30 & +22 01 52.8 &  0.007 &$-$0.198& GD153     & 4   & DA1   & 8  & 13.42 &$-$0.25&    -  &  8\\
133 & 13 15 52.80 & +46 06 36.9 &      - &      - & P172-E    & 12  & G9    & *  & 13.73 &  0.69 &    -  & 12\\
 23 & 13 41 43.57 & +28 29 49.5 &      - &      - & M3-VZ193  & 4   & G8III & *  & 14.75 &  0.90 &    -  & 10\\
134 & 14 28 43.37 & +33 10 41.5 &$-$0.354&$-$0.720& LHS 2924&13 (Note 1)&M9V& 14 & 19.58 &   -   &  4.37 & 14 \\
135 & 14 40 58.04 &$-$00 27 46.6&      - &      - & S867-V    & 12  & G5    & *  & 13.34 &  0.68 &   -   & 12\\
136 & 14 59 32.05 &$-$00 06 17.0&      - &      - & S868-G    & 12  & K2    & *  & 14.68 &  1.12 &   -   & 12\\
137 & 16 26 42.72 & +05 52 20.3 &      - &      - & P565-C    & 12  & G1    & *  & 13.34 &  0.54 &   -   & 12\\
138 & 16 28 06.72 & +34 58 48.3 &$-$0.012&  0.005 & P275-A    & 12  & A1    & *  & 10.45 &  0.00 &   -   & 12\\
139 & 16 33 52.96 & +54 28 22.1 &      - &      - & P137-F    & 12  & K1    & *  & 14.61 &  1.05 &   -   & 12\\
    &             &             &        &        &           &     &       &    &       &       &       &   \\
 27 & 16 40 41.56 & +36 21 12.4 &      - &      - &       -   &Note 2&G8IV/V& *  & 14.8: &  0.7: &   -   &   \\
140 & 17 13 22.65 &$-$18 53 33.8&      - &      - & S587-T    & 12  & G9    & *  & 12.29 &  0.72 &   -   & 12\\
141 & 17 48 58.87 & +23 17 43.7 &      - &      - & P389-D    & 12  & G2    & *  & 12.51 &  0.78 &   -   & 12\\
 35 & 18 27 13.52 & +04 03 09.4 &      - &      - &       -   &Note 3& K0   & *  & 10.0: &  0.8: &   -   & - \\
142 & 18 29 51.26 & +01 13 19.0 &      - &      - & Ser-EC51  & 17  &   -   & -  &   -   &   -   &   -   & - \\
143 & 18 29 53.79 & +01 13 29.9 &      - &      - & Ser-EC68  & 17  &   -   & -  &   -   &   -   &   -   & - \\
144 & 18 29 56.90 & +01 12 47.1 &      - &      - & Ser-EC84  & 17  &   -   & -  &   -   &   -   &   -   & - \\
145 & 18 30 05.81 & +01 13 47.3 &      - &      - & Ser-EC160 & 17  &   -   & -  &   -   &   -   &   -   & - \\
146 & 18 54 04.01 & +37 07 18.6 &      - &      - & P280-U    & 12  & K1    & *  & 12.67 &  0.99 &   -   & 12\\
147 & 19 01 55.27 & +42 29 19.6 &$-$0.001&  0.000 & P230-A(Note 4)&12& A0   & 21 & 10.02 &  0.07 &   -   & 12\\
    &             &             &        &        &           &     &       &    &    -  &       &       &   \\
148 & 19 41 23.41 &$-$03 50 56.9&$-$0.001&$-$0.004& S810-A(Note 4)&12& A0   & 21 &  9.69 &$-$0.07&   -   & 12\\
149 & 20 00 39.25 & +29 58 40.0 &  0.005 &$-$0.004& P338-C(Note 4)&12& B7.5V& 22 & 10.41 &  0.01 &   -   & 22\\
150 & 20 36 08.44 & +49 38 23.5 &  0.008 &  0.009 & CMC 513807& 21  & G0    & 21 & 10.9: &  0.4  &   -   & 11\\
 34 & 20 42 34.73 &$-$20 04 34.8&  0.355 &$-$0.098& EG141     & 4   & DA3   & 8  & 12.34 &$-$0.07&   -   &  8\\
151 & 21 04 14.75 & +30 30 21.2 &      - &      - & P340-H    & 12  & G2    & *  & 13.53 &  0.68 &   -   & 12\\
 29 & 21 52 25.36 & +02 23 20.7 &  0.023 &$-$0.303& G93-48    & 1,2 & DA3   & 8  & 12.74 &$-$0.01&$-$0.19&  9\\
152 & 22 27 16.12 & +19 16 59.2 &      - &      - & p460-E    & 12  & K1    & *  & 13.57 &  1.03 &   -   & 12\\
 30 & 22 41 44.72 & +01 12 36.5 &      - &      - & SA114-750 & 1   & B9    & 5  & 11.92 &$-$0.04&  0.12 &  9\\
153 & 23 02 32.07 &$-$03 58 53.1&  0.018 &  0.000 & S820-E    & 12  & K2    &    & 13.96 &  1.21 &   -   & 12\\
 31 & 23 12 21.60 & +10 47 04.1 &  0.130 &$-$0.011& GD246     & 1,2 & DA1   & 8  & 13.09 &$-$0.32&$-$0.33&  9\\
    &             &             &        &        &           &     &       &    &       &       &       &   \\
 32 & 23 16 12.37 &$-$01 50 34.6&      - &      - & Feige 108 & 1,2 & DAs   & 7  & 12.96 &$-$0.24&$-$0.24&  9\\
154 & 23 18 10.08 & +00 32 55.6 &      - &      - & S893-D    & 12  & G0    & *  & 12.53 &  0.65 &   -   & 12\\
155 & 23 49 47.82 & +34 13 05.1 &      - &        & CMC 516589& 11  & K5    & 21 & 12.1: &   -   &   -   & 11\\
\end{tabular}

\vspace*{2mm}

* MK spectral type estimated from $J-K$ colours ({\it via} Koorneef 1983a, 
assuming luminosity class V) or position in cluster CMD, where known.\\

\vspace*{2mm}

References in Table 1.\\

\vspace*{2mm}

1. Landolt, 1983; 2. Turnshek {\it et al.} 1990; 3. Eggen \& Sandage,
1964; 4. B. Zuckerman, 1990 personal communication; 5. Drilling \&
Landolt, 1979; 6. Klemola, 1962; 7. Greenstein, 1966; 8. McCook \& Sion,
1987; 9. Landolt, 1992; 10. Sandage \& Katem 1982; 11. Carlsberg Meridian
Catalogue 1989; 12. Lasker {\it et al.} 1988; 13.  Luyten 1979; 14.
Leggett 1992; 15. Leggett \& Hawkins 1988; 16. Leggett {\it et al.} 1994;
17. Eiroa \& Casali 1992; 18. Goodman {\it et al.} 1992; 19. Pesch 1967;
20. Chevalier \& Ilovaisky, 1991; 21. Henry Draper catalog; 22. Straizys
\& Kalytis, 1981.

\vspace*{3mm} 

Notes:\\ 

(1) FS 129 (= LHS 2397a) and FS 134 (= LHS 2924) are listed by Martin {\it
et al} (1996) amongst very cool dwarfs showing semi-regular variability
(probably from surface features and rotation), with respective ranges at I
of 0.$^m$08 and 0.$^m$06. These are considerably larger than the ranges
seen during the present observations, especially of LHS 2924 (see Table 4).

(2) FS 27 is 13.\arcmin5 from M13; its colour and magnitude are consistent
with it being an outlying subgiant member of the cluster. Its optical
magnitude and colour are inferred from the IR.

(3) FS 35 was intended to be G21-15, but the coordinates used (from
Turnshek {\it et al.} 1990) were incorrect. (Those in Landolt 1983 \& 1992
are correct, however.) This was not noticed until an anonymous star close
to the erroneous coordinates had been well observed and has accordingly
been retained here; 29 $JHK$ observations on 15 nights show no obvious
signs of variability. Its optical magnitude and colour are inferred from
the IR.

(4) FS 123 = P486-R = M67-F81; FS 147 = P230A = SAO 48018; FS 148 = S810-A
= HD 185879; FS 149 = P338-C = HD 333240.

\end{flushleft}
\end{table*}

\newpage

\begin{table*} 

\caption[]{Transmission curves of UKIRT (IRCAM) $JHK$ filters at
77K, with and without atmospheric transmission effects of a 1.2mm H$_2$O
column.}

\begin{flushleft}

\begin{tabular}{rrrrrrrrr} 

\multicolumn{3}{c}{\it J} & \multicolumn{3}{c}{\it H} & \multicolumn{3}{c}{\it 
K}\\

\multicolumn{1}{c}{$\lambda$} & \multicolumn{1}{c}{Trans.} & 
\multicolumn{1}{c}{with}&
\multicolumn{1}{c}{$\lambda$} & \multicolumn{1}{c}{Trans.} 
&\multicolumn{1}{c}{with}&
\multicolumn{1}{c}{$\lambda$} & \multicolumn{1}{c}{Trans.} 
&\multicolumn{1}{c}{with}\\

\multicolumn{1}{c}{$\mu$m} & \multicolumn{1}{c}{\%} & \multicolumn{1}{c}{atm.\%} 
&
\multicolumn{1}{c}{$\mu$m} & \multicolumn{1}{c}{\%} & \multicolumn{1}{c}{atm. 
\%} &
\multicolumn{1}{c}{$\mu$m} & \multicolumn{1}{c}{\%} & \multicolumn{1}{c}{atm. 
\%}\\

       &        &        &        &       &       \\
 1.05  &  0.000	&  0.000 & 1.42  &  0.002 & 0.001 & 1.86  &  0.000 &  0.000\\
 1.06  &  0.000	&  0.000 & 1.43  &  0.000 & 0.000 & 1.88  &  0.002 &  0.000\\
 1.07  &  0.011	&  0.011 & 1.44  &  0.001 & 0.001 & 1.90  &  0.005 &  0.001\\
 1.08  &  0.015	&  0.015 & 1.45  &  0.005 & 0.004 & 1.92  &  0.016 &  0.006\\
 1.09  &  0.028	&  0.028 & 1.46  &  0.009 & 0.008 & 1.94  &  0.059 &  0.037\\
 1.10  &  0.045	&  0.044 & 1.47  &  0.012 & 0.009 & 1.96  &  0.152 &  0.107\\
 1.11  &  0.088	&  0.085 & 1.48  &  0.022 & 0.018 & 1.98  &  0.276 &  0.253\\
 1.12  &  0.174	&  0.139 & 1.49  &  0.039 & 0.037 & 2.00  &  0.420 &  0.265\\
 1.13  &  0.334	&  0.283 & 1.50  &  0.059 & 0.060 & 2.02  &  0.552 &  0.405\\
 1.14  &  0.511	&  0.464 & 1.51  &  0.100 & 0.098 & 2.04  &  0.622 &  0.599\\
 1.15  &  0.608	&  0.529 & 1.52  &  0.176 & 0.175 & 2.06  &  0.636 &  0.556\\
 1.16  &  0.661	&  0.634 & 1.53  &  0.298 & 0.298 & 2.08  &  0.651 &  0.641\\
 1.17  &  0.705	&  0.698 & 1.54  &  0.476 & 0.474 & 2.10  &  0.682 &  0.675\\
 1.18  &  0.745	&  0.736 & 1.55  &  0.630 & 0.630 & 2.12  &  0.713 &  0.708\\
 1.19  &  0.759	&  0.751 & 1.56  &  0.713 & 0.713 & 2.14  &  0.723 &  0.723\\
 1.20  &  0.743	&  0.735 & 1.57  &  0.750 & 0.735 & 2.16  &  0.707 &  0.705\\
 1.21  &  0.727	&  0.720 & 1.58  &  0.766 & 0.749 & 2.18  &  0.686 &  0.685\\
 1.22  &  0.743	&  0.739 & 1.59  &  0.780 & 0.780 & 2.20  &  0.683 &  0.680\\
 1.23  &  0.765	&  0.765 & 1.60  &  0.811 & 0.797 & 2.22  &  0.702 &  0.702\\
 1.24  &  0.787	&  0.787 & 1.61  &  0.830 & 0.809 & 2.24  &  0.737 &  0.737\\
 1.25  &  0.784	&  0.784 & 1.62  &  0.842 & 0.842 & 2.26  &  0.771 &  0.767\\
 1.26  &  0.776	&  0.768 & 1.63  &  0.838 & 0.838 & 2.28  &  0.777 &  0.776\\
 1.27  &  0.768	&  0.733 & 1.64  &  0.832 & 0.829 & 2.30  &  0.754 &  0.754\\
 1.28  &  0.770	&  0.767 & 1.65  &  0.830 & 0.825 & 2.32  &  0.737 &  0.735\\
 1.29  &  0.795	&  0.792 & 1.66  &  0.830 & 0.829 & 2.34  &  0.739 &  0.732\\
 1.30  &  0.809	&  0.800 & 1.67  &  0.821 & 0.821 & 2.36  &  0.733 &  0.710\\
 1.31  &  0.804	&  0.775 & 1.68  &  0.831 & 0.830 & 2.38  &  0.654 &  0.641\\
 1.32  &  0.802	&  0.750 & 1.69  &  0.844 & 0.842 & 2.40  &  0.470 &  0.458\\
 1.33  &  0.805	&  0.722 & 1.70  &  0.843 & 0.842 & 2.42  &  0.268 &  0.243\\
 1.34  &  0.818	&  0.741 & 1.71  &  0.848 & 0.846 & 2.44  &  0.128 &  0.128\\
 1.35  &  0.832	&  0.568 & 1.72  &  0.851 & 0.848 & 2.46  &  0.055 &  0.054\\
 1.36  &  0.829	&  0.222 & 1.73  &  0.850 & 0.847 & 2.48  &  0.023 &  0.019\\
 1.37  &  0.818	&  0.265 & 1.74  &  0.849 & 0.839 & 2.50  &  0.010 &  0.009\\
 1.38  &  0.811	&  0.294 & 1.75  &  0.838 & 0.832 & 2.52  &  0.004 &  0.003\\
 1.39  &  0.813	&  0.498 & 1.76  &  0.846 & 0.838 & 2.54  &  0.001 &  0.000\\
 1.40  &  0.768	&  0.290 & 1.77  &  0.838 & 0.821 & 2.56  &  0.000 &  0.000\\
 1.41  &  0.602	&  0.334 & 1.78  &  0.826 & 0.786 &       &	  &\\
 1.42  &  0.320	&  0.199 & 1.79  &  0.774 & 0.737 &	  &       &\\
 1.43  &  0.124	&  0.095 & 1.80  &  0.629 & 0.483 &	  &	 & \\
 1.44  &  0.054	&  0.041 & 1.81  &  0.413 & 0.309 &	  &      & \\
 1.45  &  0.022	&  0.016 & 1.82  &  0.206 & 0.099 &	  &      & \\
 1.46  &  0.014	&  0.012 & 1.83  &  0.097 & 0.040 &	  &      & \\
 1.47  &  0.001	&  0.001 & 1.84  &  0.053 & 0.014 &	  &      & \\
 1.48  &  0.002	&  0.002 & 1.85  &  0.022 & 0.008 &	  &      & \\
 1.49  &  0.001	&  0.001 & 1.86  &  0.009 & 0.002 &	  &      & \\
 1.50  &  0.000	&  0.000 & 1.87  &  0.001 & 0.000 &	  &      & \\
       &        &        & 1.88  &  0.000 & 0.001 &	  &      & \\
		
\end{tabular}

\end{flushleft}

\end{table*}

\begin{table*} 
\caption[]{Reflectance curves of fore-optics of the IRCAM3
imager.}

\begin{flushleft}

\begin{tabular}{cccc} 

\multicolumn{1}{l}{$\lambda$($\mu$m)} & \multicolumn{1}{c}{Gold x 2} &
\multicolumn{1}{c}{Gold Dichroic} & \multicolumn{1}{c}{Zeiss ``Ag-di"}\\
      & \multicolumn{1}{c}{(generic)} &
\multicolumn{1}{c}{$\sim$40 nm
thick} & \multicolumn{1}{c}{dichroic}\\

0.7 & 0.921 & 0.876 & - \\
0.8 & 0.948 & 0.906 & 0.720\\
0.9 & 0.956 & 0.927 & 0.825\\
1.0 & 0.964 & 0.940 & 0.883\\
1.5 & 0.966 & 0.952 & 0.952\\
2.0 & 0.970 & 0.970 & 0.960\\
2.5 & 0.970 & 0.972 & 0.970\\

\end{tabular}

\end{flushleft}

\end{table*}

\newpage

\begin{table*} 

\caption[]{Positions, $K$ magnitudes and $JHK$ colours and
their uncertainties, and the number of nights observed, for the 
UKIRT Faint Standards Fundamental and Extended Lists.}

\begin{flushleft}

\begin{tabular}{lrrrrrrrr} 

\multicolumn{1}{l}{Name} & \multicolumn{2}{c}{RA \hspace{2mm}(J2000) 
\hspace{2mm} Dec} & \multicolumn{1}{c}{\it K} &
 & \multicolumn{1}{c}{$J-H$} & & \multicolumn{1}{c}{$H-K$} & \\
       &             &             &                &      &                &      &                &     \\
FS 101 & 00 13 43.58 & +30 37 59.9 & 10.384 (0.006) & [ 8] &  0.139 (0.004) & [ 8] &  0.032 (0.003) & [ 8]\\
FS 102 & 00 24 28.50 & +07 49 00.1 & 11.218 (0.005) & [ 6] &  0.318 (0.005) & [ 7] &  0.052 (0.002) & [ 6]\\
FS 1   & 00 33 54.48 &$-$12 07 58.1& 12.964 (0.004) & [ 9] &  0.387 (0.003) & [10] &  0.057 (0.004) & [ 9]\\
FS 103 & 00 36 29.60 & +37 42 54.3 & 11.731 (0.006) & [ 6] &  0.515 (0.003) & [ 6] &  0.084 (0.006) & [ 6]\\
FS 2   & 00 55 09.93 & +00 43 13.1 & 10.472 (0.004) & [12] &  0.206 (0.003) & [11] &  0.035 (0.002) & [12]\\
FS 3   & 01 04 21.63 & +04 13 36.0 & 12.823 (0.003) & [ 7] &$-$0.111 (0.003)& [ 7] &$-$0.089 (0.005) & [ 8]\\
FS 104 & 01 04 59.43 & +41 06 30.8 & 10.409 (0.005) & [ 7] &  0.101 (0.004) & [ 7] &  0.026 (0.002) & [ 7]\\
FS 105 & 01 19 08.19 & +07 34 11.5 & 10.970 (0.004) & [ 6] &  0.471 (0.008) & [ 6] &  0.080 (0.001) & [ 6]\\
FS 106 & 01 49 46.94 & +48 37 53.2 & 11.759 (0.005) & [ 5] &  0.616 (0.005) & [ 6] &  0.103 (0.006) & [ 5]\\
FS 107 & 01 54 10.14 & +45 50 38.0 & 10.231 (0.015) & [ 3] &  0.213 (0.010) & [ 3] &  0.050 (0.004) & [ 3]\\
       &             &             &                &      &                &      &                &     \\
FS 5   & 01 54 34.65 &$-$06 46 00.4& 12.339 (0.005) & [ 8] &  0.004 (0.002) & [ 7] &$-$0.005 (0.002) & [ 8]\\
FS 4   & 01 54 37.70 & +00 43 00.5 & 10.284 (0.002) & [11] &  0.239 (0.004) & [11] &  0.032 (0.002) & [11]\\
FS 6   & 02 30 16.64 & +05 15 51.1 & 13.382 (0.005) & [ 8] &$-$0.059 (0.005)& [ 8] &$-$0.069 (0.005) & [ 8]\\
FS 7   & 02 57 21.21 & +00 18 38.2 & 10.945 (0.002) & [14] &  0.126 (0.002) & [14] &  0.032 (0.001) & [14]\\
FS 108 & 03 01 09.85 & +46 58 47.7 &  9.731 (0.004) & [ 5] &  0.290 (0.004) & [ 5] &  0.059 (0.003) & [ 6]\\
FS 109 & 03 13 24.16 & +18 49 38.4 & 10.807 (0.006) & [ 6] &  0.464 (0.005) & [ 7] &  0.161 (0.004) & [ 7]\\
FS 110 & 03 41 02.22 & +06 56 15.9 & 11.324 (0.003) & [ 6] &  0.308 (0.002) & [ 7] &  0.078 (0.004) & [ 7]\\
FS 111 & 03 41 08.55 & +33 09 35.5 & 10.289 (0.006) & [ 7] &  0.262 (0.006) & [ 7] &  0.089 (0.002) & [ 7]\\
FS 112 & 03 47 40.70 &$-$15 13 14.4& 10.899 (0.005) & [ 6] &  0.264
(0.004) & [ 6] &  0.047 (0.005) & [ 6]\\
FS 10  & 03 48 50.20 &$-$00 58 31.2& 14.983 (0.011) & [ 9] &$-$0.104 (0.006)& [ 9] &$-$0.118 (0.010)& [ 9]\\
       &             &             &                &      &                &      &                &     \\
FS 113 & 04 00 14.07 & +53 10 38.5 & 12.432 (0.005) & [ 6] &  0.372 (0.003) & [ 6] &  0.109 (0.005) & [ 6]\\
FS 114 & 04 19 41.72 & +16 45 22.4 & 13.445 (0.006) & [ 5] &  0.604 (0.002) & [ 6] &  0.337 (0.006) & [ 5]\\
FS 115 & 04 23 18.17 & +26 41 16.4 & 10.171 (0.004) & [ 6] &  1.585 (0.009) & [ 7] &  0.772 (0.004) & [ 7]\\
FS 116 & 04 23 50.18 & +26 40 07.7 & 10.949 (0.002) & [ 7] &  1.261 (0.007) & [ 6] &  0.511 (0.003) & [ 7]\\
FS 117 & 04 23 56.61 & +26 36 38.0 & 10.078 (0.004) & [ 7] &  0.950 (0.008) & [ 7] &  0.437 (0.002) & [ 7]\\
FS 118 & 04 24 33.49 & +26 33 37.8 & 10.446 (0.005) & [ 6] &  0.723 (0.005) & [ 6] &  0.266 (0.001) & [ 6]\\
FS 11  & 04 52 58.92 &$-$00 14 41.6& 11.254 (0.002) & [18] &  0.065 (0.001) & [18] &  0.022 (0.001) & [18]\\
FS 119 & 05 02 57.44 &$-$01 46 42.6&  9.851 (0.004) & [ 6] &  0.045 (0.006) & [ 6] &  0.019 (0.002) & [ 6]\\
FS 12  & 05 52 27.66 & +15 53 14.3 & 13.916 (0.006) & [13] &$-$0.115 (0.003)& [13] &$-$0.094 (0.005)& [13]\\
FS 13  & 05 57 07.59 & +00 01 11.4 & 10.140 (0.002) & [12] &  0.313 (0.002) & [13] &  0.048 (0.001) & [12]\\
       &             &             &                &      &                &      &                &     \\
FS 120 & 06 14 01.44 & +15 09 58.3 & 10.626 (0.006) & [ 6] &  0.503 (0.004) & [ 7] &  0.199 (0.006) & [ 7]\\
FS 121 & 06 59 46.82 &$-$04 54 33.2& 11.315 (0.003) & [ 7] &  0.561 (0.003) & [ 7] &  0.108 (0.003) & [ 7]\\
FS 122 & 07 00 52.02 & +48 29 24.0 & 11.338 (0.003) & [ 7] &  0.268 (0.003) & [ 6] &  0.046 (0.003) & [ 7]\\
FS 14  & 07 24 14.40 &$-$00 33 04.1& 14.198 (0.008) & [ 8] &$-$0.065 (0.007)& [ 8] &$-$0.031 (0.010)& [ 8]\\
FS 15  & 08 51 05.81 & +11 43 46.9 & 12.348 (0.003) & [13] &  0.337 (0.003) & [13] &  0.054 (0.002) & [13]\\
FS 123 & 08 51 11.88 & +11 45 21.5 & 10.211 (0.004) & [10] &$-$0.012 (0.002)& [10] &$-$0.022 (0.004)& [10]\\
FS 16  & 08 51 15.01 & +11 49 21.2 & 12.628 (0.004) & [11] &  0.290 (0.003) & [11] &  0.041 (0.002) & [11]\\
FS 17  & 08 51 19.31 & +11 52 10.4 & 12.274 (0.002) & [10] &  0.327 (0.003) & [10] &  0.056 (0.002) & [10]\\
FS 18  & 08 53 35.51 &$-$00 36 41.7& 10.527 (0.002) & [11] &  0.252 (0.003) & [11] &  0.041 (0.002) & [11]\\
FS 124 & 08 54 12.07 &$-$08 04 58.9& 10.776 (0.004) & [ 6] &  0.481 (0.006) & [ 7] &  0.266 (0.005) & [ 7]\\
       &             &             &                &      &                &      &                &     \\
FS 125 & 09 03 20.60 & +34 21 03.9 & 10.374 (0.005) & [10] &  0.376 (0.002) & [ 9] &  0.052 (0.003) & [10]\\
FS 126 & 09 19 18.73 & +10 55 54.2 & 11.680 (0.005) & [10] &  0.564 (0.003) & [10] &  0.105 (0.003) & [ 9]\\
FS 127 & 10 06 29.03 & +41 01 26.6 & 11.654 (0.004) & [10] &  0.261 (0.004) & [ 9] &  0.040 (0.003) & [10]\\
FS 19  & 10 33 42.75 &$-$11 41 38.3& 13.782 (0.006) & [ 9] &$-$0.085 (0.005)& [ 9] &$-$0.115 (0.007)& [ 9]\\
FS 128 & 11 05 10.40 & +07 06 48.7 & 12.078 (0.004) & [ 5] &  0.616 (0.003) & [ 5] &  0.309 (0.005) & [ 6]\\
FS 20  & 11 07 59.93 &$-$05 09 26.1& 13.501 (0.010) & [ 6] &$-$0.038 (0.006)& [ 6] &$-$0.065 (0.008)& [ 6]\\
FS 129 & 11 21 48.95 &$-$13 13 07.9& 10.709 (0.006) & [ 8] &  0.724 (0.006) & [ 8] &  0.425 (0.002) & [ 7]\\
FS 130 & 11 24 55.92 & +34 44 38.5 & 12.252 (0.003) & [ 6] &  0.596 (0.006) & [ 7] &  0.099 (0.004) & [ 6]\\
FS 21  & 11 37 05.15 & +29 47 58.4 & 13.147 (0.002) & [ 9] &$-$0.069 (0.004)& [ 9] &$-$0.090 (0.004)& [ 9]\\
FS 131 & 12 14 25.40 & +35 35 55.6 & 11.314 (0.005) & [ 5] &  0.259 (0.003) & [ 6] &  0.032 (0.004) & [ 5]\\

\end{tabular}
\end{flushleft}

\end{table*}

\begin{table*}

\contcaption{}

\begin{flushleft}
\begin{tabular}{lrrrrrrrr} 
\multicolumn{1}{l}{Name} & \multicolumn{2}{c}{RA \hspace{2mm}(J2000) 
\hspace{2mm} Dec} & \multicolumn{1}{c}{\it K} &
 & \multicolumn{1}{c}{\it J-H} & & \multicolumn{1}{c}{\it H-K} & \\
       &             &             &                &      &                &      &                &     \\
FS 132 & 12 21 39.36 &$-$00 07 13.3& 11.836 (0.004) & [ 6] &  0.290 (0.003) & [ 7] &  0.034 (0.005) & [ 6]\\
FS 33  & 12 57 02.30 & +22 01 52.8 & 14.254 (0.005) & [ 6] &$-$0.112 (0.002)& [ 5] &$-$0.103 (0.008)& [ 5]\\
FS 133 & 13 15 52.80 & +46 06 36.9 & 11.877 (0.004) & [ 5] &  0.367 (0.005) & [ 5] &  0.060 (0.003) & [ 4]\\
FS 23  & 13 41 43.57 & +28 29 49.5 & 12.375 (0.003) & [11] &  0.527 (0.003) & [11] &  0.066 (0.003) & [11]\\
FS 134 & 14 28 43.37 & +33 10 41.5 & 10.747 (0.004) & [ 6] &  0.746 (0.004) & [ 6] &  0.424 (0.002) & [ 6]\\
FS 135 & 14 40 58.04 &$-$00 27 46.6& 11.594 (0.003) & [ 5] &  0.344 (0.003) & [ 5] &  0.046 (0.002) & [ 6]\\
FS 136 & 14 59 32.05 &$-$00 06 17.0& 11.883 (0.005) & [ 6] &  0.553 (0.004) & [ 6] &  0.090 (0.004) & [ 6]\\
FS 137 & 16 26 42.72 & +05 52 20.3 & 11.829 (0.006) & [ 5] &  0.273 (0.006) & [ 7] &  0.045 (0.003) & [ 6]\\
FS 138 & 16 28 06.72 & +34 58 48.3 & 10.412 (0.006) & [ 7] &  0.030 (0.006) & [ 6] &  0.002 (0.002) & [ 7]\\
FS 139 & 16 33 52.96 & +54 28 22.1 & 12.103 (0.006) & [ 7] &  0.512 (0.006) & [ 7] &  0.091 (0.003) & [ 7]\\
       &             &             &                &      &                &      &                &     \\
FS 27  & 16 40 41.56 & +36 21 12.4 & 13.128 (0.005) & [10] &  0.306 (0.005) & [10] &  0.048 (0.004) & [10]\\
FS 140 & 17 13 22.65 &$-$18 53 33.8& 10.369 (0.006) & [ 5] &  0.369 (0.004) & [ 5] &  0.060 (0.005) & [ 5]\\
FS 141 & 17 48 58.87 & +23 17 43.7 & 10.812 (0.006) & [ 7] &  0.303 (0.004) & [ 7] &  0.061 (0.003) & [ 7]\\
FS 35  & 18 27 13.52 & +04 03 09.4 & 11.748 (0.005) & [14] &  0.369 (0.003) & [15] &  0.084 (0.002) & [15]\\
FS 142 & 18 29 51.26 & +01 13 19.0 & 13.393 (0.011) & [ 2] &  1.495 (0.005) & [ 2] &  0.750 (0.006) & [ 2]\\
FS 143 & 18 29 53.79 & +01 13 29.9 & 13.018 (0.010) & [ 4] &  2.336 (0.011) & [ 3] &  1.200 (0.005) & [ 4]\\
FS 144 & 18 29 56.90 & +01 12 47.1 & 11.046 (0.001) & [ 2] &  2.513 (0.017) & [ 2] &  1.302 (0.010) & [ 2]\\
FS 145 & 18 30 05.81 & +01 13 47.3 & 12.160 (0.006) & [ 4] &  3.234 (0.049) & [ 4] &  1.562 (0.009) & [ 4]\\
FS 146 & 18 54 04.01 & +37 07 18.6 & 10.141 (0.005) & [ 5] &  0.511 (0.005) & [ 5] &  0.070 (0.003) & [ 5]\\
FS 147 & 19 01 55.27 & +42 29 19.6 &  9.866 (0.004) & [ 4] &  0.028 (0.004) & [ 5] &  0.015 (0.003) & [ 5]\\
       &             &             &                &      &                &      &                &     \\
FS 148 & 19 41 23.52 &$-$03 50 56.1&  9.478 (0.004) & [ 5] &  0.020 (0.008) & [ 5] &  0.011 (0.005) & [ 5]\\
FS 149 & 20 00 39.25 & +29 58 40.0 & 10.086 (0.009) & [ 6] &  0.023 (0.008) & [ 6] &$-$0.003 (0.003)& [ 6]\\
FS 150 & 20 36 08.44 & +49 38 23.5 &  9.960 (0.005) & [ 8] &  0.154 (0.007) & [ 9] &  0.044 (0.003) & [ 9]\\
FS 34  & 20 42 34.73 &$-$20 04 34.8& 13.000 (0.004) & [ 9] &$-$0.077 (0.003)& [ 9] &$-$0.074 (0.005)& [ 9]\\
FS 151 & 21 04 14.75 & +30 30 21.2 & 11.876 (0.007) & [ 8] &  0.274 (0.006) & [ 8] &  0.061 (0.005) & [ 8]\\
FS 29  & 21 52 25.36 & +02 23 20.7 & 13.311 (0.003) & [11] &$-$0.068 (0.002)& [11] &$-$0.070 (0.003)& [11]\\
FS 152 & 22 27 16.12 & +19 16 59.2 & 11.057 (0.005) & [ 9] &  0.521 (0.005) & [ 8] &  0.062 (0.004) & [ 9]\\
FS 30  & 22 41 44.72 & +01 12 36.5 & 12.022 (0.003) & [15] &$-$0.042 (0.002)& [13] &$-$0.031 (0.002)& [14]\\
FS 153 & 23 02 32.07 &$-$03 58 53.1& 10.896 (0.004) & [ 6] &  0.580 (0.005) & [ 7] &  0.105 (0.002) & [ 7]\\
FS 31  & 23 12 21.60 & +10 47 04.1 & 14.037 (0.007) & [10] &$-$0.130 (0.005)& [10] &$-$0.099 (0.005)& [10]\\
       &             &             &                &      &                &      &                &     \\
FS 32  & 23 16 12.37 &$-$01 50 34.6& 13.676 (0.007) & [ 8] &$-$0.107 (0.003)& [ 8] &$-$0.086 (0.005)& [ 8]\\
FS 154 & 23 18 10.08 & +00 32 55.6 & 11.064 (0.005) & [ 6] &  0.264 (0.003) & [ 7] &  0.042 (0.003) & [ 7]\\
FS 155 & 23 49 47.82 & +34 13 05.1 &  9.413 (0.006) & [ 6] &  0.511 (0.003) & [ 5] &  0.085 (0.003) & [ 6]\\
\end{tabular}

\vspace*{2mm}

Notes to Table 4.\\

\vspace*{2mm}

(1) The standard error of the mean is shown in parentheses.\\

(2) The number of nights from which IRCAM3 data have been retained are
shown in square brackets.\\

(3) FS 18 is a double star with separation = 1{\arcsec}.36, p.a.
96\degree, $\Delta$(K) = 2.$^m$2 (Y. Clenet 1999, personal
communication).\\

\end{flushleft}

\end{table*}


\begin{thebibliography}{99}

%
\bibitem{} Allen, D.A. \& Cragg, T.A., 1983. MNRAS, 203, 777

\bibitem{} Bersanelli, M., Bouchet, P. \& Falomo, R., 1991. A \& A, 252,
854

\bibitem{} Bessell, M.S. \& Brett, J.M., 1988. PASP, 100, 1134

\bibitem{} Blackwell, D.E., Leggett, S.K., Petford, A.D., Mountain, C.M.
\& Selby, M.J., 1983. MNRAS, 205, 897.

\bibitem{} Bouchet, P., Manfroid, J. \& Schmider, F.X., 1991. A \& A Supp.,
91, 409

\bibitem{} Carrasco, L., Recillas-Cruz, E., Garcia-Barreto, A.,
Cruz-Gonzalez, I. \& Serrano, A.P.G., 1991. PASP, 103, 987

\bibitem{} Carter, B.S., 1990. MNRAS, 242,1

\bibitem{} Carter, B.S. \& Meadows, V.S., 1995. MNRAS, 276, 734

\bibitem{} Casali, M.M \& Hawarden, T.G., 1992. The JCMT-UKIRT
Newsletter, 4, 35

\bibitem{} Carlsberg Meridian Catalogue La Palma, No 4, 1989. Copenhagen
University observatory, Royal Greenwich Observatory \& Real Instituto y
Observatorio de la Armarda en San Fernando (Servico Publicaciones
Armarda).

\bibitem{} Chapman, R., Beard, S.M., Mountain, C.M., Pettie, D.G., Pickup,
D.A. \& Wade, R., 1990. Proc. SPIE, 1235, 34

\bibitem{} Chen, H., Tokunaga, A.T., Strom, K.M. \& Hodapp, K.-W., 1993.
ApJ, 407, 639

\bibitem{} Chevalier, C. \& Ilovaisky, S.A., 1991. A \& A Suppl., 90, 225

\bibitem{} Cohen, M., Walker, R.G., Carter, B., Hammersley, P.L., Kidger,
M.R. \& Noguchi, K., 1999. AJ, 117, 1864

\bibitem{} Drilling, J.S. \& Landolt, A.U., 1979. ApJ, 84, 783

\bibitem{} Eggen, O.J. \& Sandage, A.R., 1964. ApJ, 140, 130

\bibitem{} Elias, J.H. Frogel, J.A., Matthews, K \& Neugebauer, G., 1982.
AJ, 87, 1029

\bibitem{} Elias, J.H., Frogel, J.A., Hyland, A.R. \& Jones, T.J., 1983.
AJ, 88, 1027

\bibitem{} Eiroa, C. \& Casali, M.M., 1992. A\&A 262, 468

\bibitem{} Forbes, J.D., 1848. Phil. Trans., 132, 225

\bibitem{} Frogel, J.A., Persson, S.E., Aaronson, M. \& Matthews, K.,
1978. ApJ, 220, 75

\bibitem{} Glass, I. S., 1974. Mon Notes Astr. Soc. S. Africa, 33, 53

\bibitem{} Goodman, A.A., Jones, T.J., Lada, E.A. \& Myers, P.C., 1992. ApJ, 
399, 108 

\bibitem{} Greenstein, J.L., 1966. ApJ, 144, 496 

\bibitem{} Guarnieri, M.D., Dixon, R.I. \& Longmore, A.J., 1991. PASP,
103, 675

\bibitem{} Hawarden, T.G., Rees, N.P., Chuter, T.C., Chrysostomou, A.C.,
Cavedoni, C.P., Rohloff, R-R., Pitz, E., Pettie, D.G., Bennett, R.J. \&
Atad-Ettedgui, E., 1999. Proc SPIE, 3785, 82

\bibitem{} Hodapp, K.-W. \& Rayner, J., 1991. AJ, 102, 1108

\bibitem{} Hunt, L.K., Mannucci, F., Testi, L., Migliorni, S., Stanga,
R.M., Baffa, C., Lisi, F. \& Vanzi, L., 1998. AJ, 115, 2594

\bibitem{} Johnson, H.L., 1996. Ann. Rev. Astron. Astrophys., 3, 193

\bibitem{} Klemola, A.R., 1962. AJ, 67, 740 

\bibitem{} Koorneef, J., 1983a. A \& A, 128, 84

\bibitem{} Koorneef, J., 1983b. A \& A Suppl., 51, 489

\bibitem{} Landolt, A.U., 1983. AJ, 88, 439

\bibitem{} Landolt, A.U., 1990, PASP, 102, 1382

\bibitem{} Landolt, A.U., 1992. AJ, 104, 340

\bibitem{} Lasker, B.M., Sturch, C.R., Lopez, C., Mallama, A.D., McLaughlin, 
S.F.,
Russell, J.F., Wisniewski, W.Z., Gillespie, B.A., Jenkner,H., Siciliano,
E.D., Kenny, D., Baumert, J.H., Goldberg,A.M., Henry, G.W., Kemper, E \&
Siegel, M.J., 1988. Ap.J. Supp, 68, 1

\bibitem{} Leggett, S.K. 1992. Ap.J. Supp., 82, 351

\bibitem{} Leggett, S.K., Harris, H.C. \& Dahn, C., 1994. AJ 108, 944

\bibitem{} Leggett, S.K. \& Hawkins, M.R.S., 1988. MNRAS 234, 1065

\bibitem{} Luyten, W.J., 1979, LHS Catalogue Minneapolis: University of 
Minnesota

\bibitem{} Martin, E.L., Zapatero-Osorio, M.R., \& Rebolo, R., 1996. 9th
Cambridge workshop on ``Cool stars, stellar systems \& the sun", eds.
R.Pallavicini \& A.Dupree, ASP Conf. Series 109, 615 

\bibitem{} McClean, I.S., Chuter, T.C., McCaugrean, M.J.\& Rayner,
J.T., 1986. Proc. SPIE, 637, 430

\bibitem{} McCook, G.P., \& Sion, E.M., 1987. ApJS, 65, 603 

\bibitem{} McGregor, P.J., 1994. PASP, 106, 508

\bibitem{} Persson, S.E., Murphy, D.C., Krzeminiski, W., Roth, M. \&
Rieke,M.J., 1998. AJ, 116, 2475

\bibitem{} Pesch, P., 1967. ApJ, 148, 781

\bibitem{} Puxley, P.J., Sylvester, J, Pickup, D.A., Paterson, M.J.,
Laird, D.C. \& Atad-Ettedgui, E, 1994. Proc. SPIE, 2198, 350 

\bibitem{} Sandage, A.R. \& Katem, B., 1982. AJ, 87, 537

\bibitem{} Simons, D. \& Tokunaga, A., 2001. PASP submitted

\bibitem{} Straizys, V. \& Kalytis, R., 1981. Act. Astron., 31, 93

\bibitem{} Tapia, M., Neri, L. \& Roth, M., 1986. Rev. Mex. Astron.
Astrof., 13, 115

\bibitem{} Tokunaga, A. \& Simons, D., 2001. PASP submitted

\bibitem{} Turnshek, D.A., Bohlin, R.L., Williamson, R.L. II, Lupie, O.L.,
Koorneef, J, \& Morgan, D.H., 1990. AJ, 99, 1243 

\bibitem{} Young, A.T., Milone, E.F. \& Stagg, C.P., 1994. A \& A Suppl.,
105, 259

\end{thebibliography}
\end{document}